\documentclass{pnastwo}

\usepackage[dvips]{graphicx}
\usepackage{amssymb,amsfonts,amsmath}

\url{www.pnas.org/cgi/doi/10.1073/pnas.0709640104}
\copyrightyear{2008}
\issuedate{Issue Date}
\volume{Volume}
\issuenumber{Issue Number}

\begin{document}

\title{Ideal Glass Transitions by Random Pinning}

\author{Chiara Cammarota\affil{1}{Institut Physique Th\'eorique (IPhT) CEA Saclay, and CNRS URA 2306, 91191 Gif Sur Yvette, France},
Giulio Biroli\affil{1}{Institut Physique Th\'eorique (IPhT) CEA Saclay, and CNRS URA 2306, 91191 Gif Sur Yvette, France}}

\contributor{Submitted to Proceedings of the National Academy of Sciences
of the United States of America}

\maketitle

\begin{article}
\begin{abstract}
We study the effect of freezing the positions of a fraction $c$ of particles 
from an equilibrium configuration of a supercooled liquid at a temperature $T$.
We show that within the random first-order transition theory pinning particles leads to an ideal glass
transition for a critical fraction $c=c_{K}(T)$ even for moderate super-cooling, {\it e.g.} close to the Mode-Coupling transition 
temperature. We first derive the phase diagram in the $T-c$ plane by mean field approximations. Then, by applying a real-space renormalization group method, we obtain the critical properties for $|c-c_{K}(T)|\rightarrow 0$, in particular the divergence of length and time scales. These are dominated by two zero-temperature fixed points. We also show that for $c=c_{K}(T)$ the typical distance between frozen particles 
 is related to the static point-to-set lengthscale of the unconstrained liquid. We discuss what are the
main differences when particles are frozen in other geometries and not from an equilibrium configuration.
Finally, we explain why the glass transition induced by freezing particles provides a new and very promising avenue of research to probe the glassy state and ascertain, or disprove, the validity of the theories of the glass transition.
\end{abstract}



\dropcap{A}lmost any liquid shows an impressive growth of the relaxation time when supercooled 
below the melting point: indeed typical time-scales increase from picoseconds to hours in a rather restricted
temperature window.  This increase is so steep that below a certain temperature, $T_g$, 
it is not possible to equilibrate the system, which then freezes in an amorphous solid, called glass \cite{revBB}. 
The experimental glass transition temperature $T_g$ is not a critical temperature at all, but only 
defines a cross-over between the system relaxation time and the observation timescale.   
A recurrent question in the literature is whether a true thermodynamic phase transition to an ideal glass 
state takes place at a temperature $T_K<T_g$. This would explain the fast growth of the relaxation time 
and several other phenomena observed experimentally. 
Other scenarii, based on purely dynamical transitions at zero (or finite) temperature or even no transition at all, 
might also provide a correct explanation \cite{gcrev,gilles}.\\
The study of the glass transition is hampered by the slowing down of the dynamics which is much more severe than for usual critical phenomena. 
Recent work on growing lengthscales \cite{oupbook} suggests that even if a transition is present the 
corresponding critical length that can be observed in experiments is no longer than a few inter-particle distances, meaning 
that experimental systems can be equilibrated only rather far from the putative transition point. 
This makes quite difficult to prove (or disprove) the existence of the transition and also to 
contrast theories solely on the basis of their critical properties.   Here we propose a way to bypass these problems: 
we show that an ideal glass transition\footnote{'Ideal glass transition' usually refers to the transition, conjectured to take place at a finite temperature $T_K$, towards an ideal glass state. We shall extend this name to all transitions induced by random pinning even though there are some important differences 
that we shall discuss in the following.} can be induced by a suitable random perturbation even for moderate 
supercooling. Moreover, we show that  
the induced ideal glass phase can be accessed and sampled easily.
Contrary to the usual situation, where the existence
(or the absence) of the ideal glass transition can only be advocated on the basis of doubtful extrapolations, one can now
approach it from both ends. As a consequence, determining whether there is indeed a true transition becomes feasible by using the usual machinery of standard phase transitions, in particular finite-size scaling. A crucial point is that this "high temperature" ideal glass transition is expected within the random first-order transition (RFOT) \cite{ktw} 
theory only. Thus, its study should lead to very sharp tests allowing to ascertain the validity of the theory or disprove it.\\
The suitable random perturbation we focus on consists in freezing the positions of an infinite subset of particles of an equilibrated configuration. 
This kind of procedure, first performed in \cite{kobparisi}, has recently been used to test the mosaic picture of the RFOT theory \cite{bbrev,bb}, show the existence of medium range amorphous order \cite{bbcgv} and study the growth of a static length \cite{bbcgv,ts}, called point-to-set \cite{ms}. 
In a very recent work Berthier and Kob have focused on several different geometries of frozen particles \cite{bk}. 
Their most remarkable finding is obtained when the positions of all particles outside two walls are frozen and when particles 
are pinned at random with a concentration $c$ (see also \cite{Kim1,Kim2,Procaccia}). In both these cases the relaxation time increases very fast
when the fraction of frozen particles is increased or the distance between the two walls is diminished, even at temperatures close to the onset of glassy dynamics. The increase is so dramatic that
it was claimed it could correspond to a true divergence \cite{bk,Kim1,Procaccia}.\\
Here we show that within the RFOT approach this is indeed the case and that this divergence is related to an ideal glass transition. We first present phenomenological arguments supporting the presence of this phenomenon. We then theoretically derive the existence of this transition and its critical properties and discuss the growth of length and time scales by several methods. We make use of mean-field techniques developed for disordered systems \cite{revAC} and of the renormalization-group approach recently introduced in \cite{cbtt}. We show that the resulting theoretical problem appears to be related to nucleation close to a first-order phase transition in the presence of quenched disorder.  
The role of disorder is, actually, crucial and will be discussed thoroughly: disorder is expected to wipe 
out the transition in the case of two walls in dimensions three and less and, in the case of randomly frozen particles, in dimensions two and less.
In the remaining cases, {\it i.e.} above the lower critical dimension, 
disorder should give rise to very unusual finite-size effects. 
Finally, we propose several ways to study these ideal glass transitions induced by freezing
particles and we discuss the main reasons that make such studies 
feasible, in contrast with the usual case of the low-temperature ideal glass 
transition. We illustrate our main ideas by presenting numerical 
simulations of the Ising model when spins are pinned at random. \\
{\bf General Arguments}.
As stressed previously, our results only hold within the RFOT theory, which 
explains the static and 
dynamic properties of supercooled liquids in terms of the competition between 
 the surface tension $\Upsilon$ and the configurational entropy density $s_c(T)$.  
The former is a measure of the extra free-energy cost paid when two different amorphous phases are in contact through a common surface. The latter quantifies the multiplicity of amorphous phases in which the liquid can freeze. In mean-field models and approximate computations \cite{mezard10} applied to
 finite dimensional liquids, $s_c$ is found to vanish at a finite temperature $T_K$. On a lengthscale $l$ (in spatial dimension $d$) the configurational entropy and surface tension respectively scale as $Ts_cl^d$ and $\Upsilon l^\theta$ (with $\theta\le d-1$). 
In consequence, on lengthscales smaller than $l_{PS}=(\Upsilon/Ts_c)^{1/(d-\theta)}$ it is preferable for the system to develop amorphous order, localizing in a given amorphous phase. Instead, for $l>l_{PS}$, it is better to break the amorphous order, pay the extra free-energy cost and gain
entropy. This gives rise to the so-called mosaic state envisioned as an assembly of patches of length $l_{PS}$ \cite{ktw,bb}: a kind of micro-phase separated state where the number of phases is actually huge (proportional to $e^{s_c l_{PS}^d}$). Recently, instantons computations confirmed the scenario depicted above, the existence of $l_{PS}$, defined as a point-to-set length, and its scaling with $Ts_c$ \cite{dzero,silvio,franzoupchapter}. \\ 
What are the consequences, within the RFOT theory, of freezing at random the positions of a fraction $c$ of particles of an 
 equilibrated configuration? 
Freezing actually means blocking the particle positions. 
The remaining particles then evolve in this frozen background.  
For this constrained liquid both the number of possible amorphous phases and the extra free-energy cost of the interfaces  between them, $\Upsilon(T,c)$, change with $c$.
The multiplicity of amorphous phases clearly is less than that 
of the unconstrained system. 
By pinning a particle one reduces the possible low free-energy configurations of the surrounding ones and, hence, diminishes the configurational entropy density by an amount $Y$ (not to be confused with $\Upsilon$).
Thus, at least for small $c$, one finds $s_c(T,c)\simeq s_c(T)-cY(T)$. 
Generically, one expects
$s_c(T,c)$ to be a monotonously decreasing function of $c$, eventually vanishing\footnote{Note that $Y(T)$ is a microscopic configurational entropy loss and is not expected to 
vanish even at high temperature.} 
 for a critical fraction of frozen particles $c_K(T)\simeq s_c(T)/Y(T) <1$. Repeating the previous 
argument about the competition between configurational entropy and surface free energy cost, one then finds that the configurations of the constrained liquid
are organized in 
a mosaic with larger tiles: the point-to-set length, $l_{PS}(T,c)=(\Upsilon(T,c)/s_c(T,c))^{1/(d-\theta)}$, increases as a function of $c-c_K$ and diverges at the point $c_K(T)$ at which $s_c(T,c)$ becomes zero. 
Thus, we find that the condition $s_c(T,c_K(T))=0$, or equivalently $s_c(T_K(c),c)=0$, defines a line of ideal glass transitions. Its starting point, at $c=0$, coincides with  the ideal glass transition of the unperturbed liquid; 
its endpoint corresponds to the temperature at which $\Upsilon(T,c)$ vanishes. Within mean-field theory this 
happens at the onset temperature $T_o$ \cite{franzoupchapter,pspinfreezing}, see the top panel of Fig.1.
By going beyond mean-field theory and by taking into account the effect of the quenched disorder induced by the pinned particles, we find (see later) that the endpoint of the
critical line actually takes place at a temperature $T_h<T_o$, see bottom panel of Fig.1. The value of $T_h$ depends on the ratio $T_d/T_K$, {\it i.e.} on the fragility of the liquid: when $T_d/T_K$ decreases (more fragile liquids), $T_h/T_d$ increases. 
Note that whether or not $T_K(0)$ is different from
zero, {\it i.e.} whether or not a finite temperature ideal glass transition exists for the unperturbed liquid, is not crucial for the present discussion. 
In conclusion, we find that ideal glass transitions can be induced at temperatures between $T_K$ and $T_h$ by increasing the fraction of frozen particles in the system\footnote{We are not interested in the regime corresponding 
to very large $c$ where other physical effects, such as percolation of the frozen particles and Lorentz-gas physics, come into play \cite{Kim1,Kim2}.}.
The case corresponding to freezing all particles in the system except those between two infinite flat walls at distance $d$ is similar and will be discussed at the end of the paper.
We end this section with two important remarks on the newly found glass transitions. 
First, pinning particles can be considered as a very nonlinear way to probe the supercooled liquid state. In
consequence, the existence of
\vspace{-0.95cm}
\begin{figure}[h]
\hspace{-0.35cm}
\includegraphics[width=0.45\textwidth]{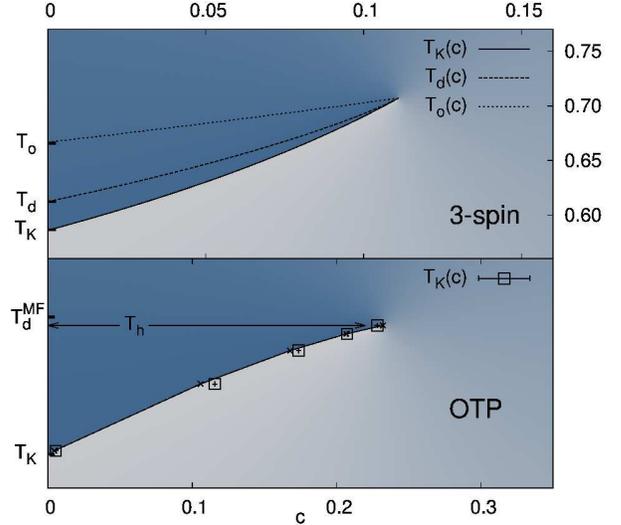}\\
\vspace{-0.4cm}
\caption{Top panel: Phase diagram for the $p=3$-spin disordered model \cite{pspinfreezing}.
The continuous line corresponds to ideal glass transitions taking place at $T_K(c)$. The two dotted lines (from top to bottom) correspond respectively to the onset temperature $T_o(c)$ and the mode coupling critical temperatures $T_d(c)$. 
 Bottom Panel: Phase diagram obtained by the renormalization group approach for $T_d/T_K\simeq 1.4$ (corresponding schematically to OTP). The variable on the x-axis is proportional to the concentration of pinned particles.
 The squares are the results obtained by RG for the ideal glass transition line $T_K(c)$. The continuous line corresponds to the phenomenological result $c_K \propto s_c/Y$.
The endpoint of the transition line takes place at a temperature $T_h<T_d$ in contrast to the mean field case.\vspace{-1.2cm}\\}
\label{fig1}
\end{figure}
$\phantom{a}$\\
$\phantom{a}$\\
a glass transition for $c=c_K(T)$ reveals the presence of a new characteristic static lengthscale $l_K(T)$ for the {\it unconstrained} liquid. This lengthscale corresponds to the critical mean distance between frozen particles: $l_K(T)=1/c_K^{1/d}$.
Since $c_K(T)\propto s_c(T)$, the new static lengthscale, $l_K$, is expected to diverge at $T=T_K$. It is interesting
to remark that the divergence is milder than the one of $l_{PS}$: $l_K\propto 1/s_c^{1/d}\propto l_{PS}^{(d-\theta)/d}$. \\
The second very important issue is that, as we shall discuss later, the phase diagram of Fig.1 
is not expected within any theoretical approach other than the RFOT one.
Thus, the new phenomena discussed in this work should allow one to test glass transition theories in a very stringent way. 
One could wonder, on the other hand, why showing the existence of ideal glass transition as a function of $c$ is not as difficult as to show it as a function of temperature. In the usual case the main problem is that the low temperature
phase is not known. In the frozen-particle case, this is not the case! 
Actually, the configuration ${\mathcal C}_{nf}$ of the nonfrozen particles just after freezing out the others is automatically an equilibrium configuration for the new system  \cite{kobparisi,franzoupchapter,krakoviack}, even for $c>c_K$. In this regime the nonfrozen particles just vibrate around the ideal glass configuration, which is now known: it corresponds to ${\mathcal C}_{nf}$ and a lump of configurations around it. 
In consequence,
one can now approach the transition from both ends (instead of making extrapolations from high temperature) and show whether there is indeed a transition, study its critical properties, devise clever algorithm to equilibrate the system, etc.  We postpone a more detailed discussion to the end of the paper. 
In the following we present theoretical arguments to corroborate the previous phenomenological ones.\\
\vspace{-0.8cm}
\begin{figure}[h]
\hspace{-0.35cm}
\includegraphics[width=0.45\textwidth]{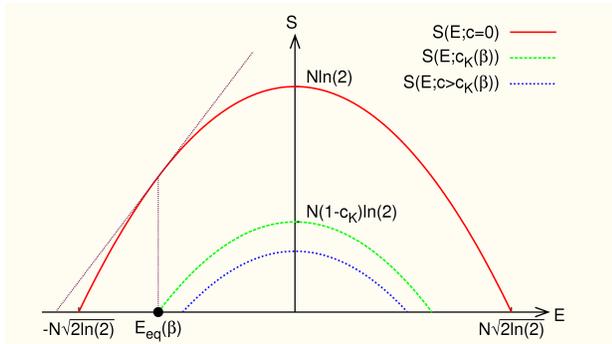}\\
\vspace{-0.2cm}
\caption{Configurational entropy for the REM as a function of $E$ for $c=0,c=c_K,c>c_K$ (from top to bottom). The equilibrium energy is stuck at the value $E_{eq}(\beta)$. For $c<c_K$ the partition function is dominated by configurations with energy 
$E_{eq}(\beta)$ (the straight line has a slope $\beta$). For $c>c_K$ the partition function is dominated by the contribution from the reference configuration; higher states lead to a sub-leading contribution.\vspace{-1.3cm}\\}\label{fig2}
\end{figure}
$\phantom{a}$\\
$\phantom{a}$\\
{\bf Phase Diagram and Mean Field Theory}.
An easy way to study within mean-field theory the effect of freezing degrees of freedom is to focus
on the Random Energy Model (REM) \cite{Rde} (see \cite{ricci} for a similar computation). The REM is the simplest mean field model displaying an entropy vanishing transition {\it \`a la} Kauzmann \cite{revBB}
and it is often used to understand some aspects of supercooled liquids. 
It is a spin model in which the energies of the configurations are independent and identically distributed (i.i.d.) Gaussian 
random variables of variance $N$, where $N$ is the number of spins (such a scaling is needed to obtain a correct
thermodynamic limit). The number of configurations with energies between $E$ and $E+dE$ reads
${\mathcal N}(E)=2^N\exp[-E^2/(2N)]/\sqrt{2\pi N}$. The equilibrium energy is given by the condition $\beta=\partial \ln {\mathcal N}(E)/\partial E$, i.e. $\beta=-E_{eq}(\beta)/N$. The entropy $S$ is simply equal to $\ln {\mathcal N}(E_{eq})$. 
The model has a transition at the temperature $\beta_K=1/T_K=\sqrt{2\ln 2}$ at which $S(E_{eq})$ vanishes. For $T<T_K$ the energy as a function of temperature does not vary any longer.
Now, take a high temperature equilibrium configuration $\alpha$ and freeze a fraction $c$ of its spins. We call this set ${\mathcal C}_f$. In order to study 
the thermodynamics of the nonfrozen spins, 
one has to sum over all configurations having the spins in ${\mathcal C}_f$ 
in the same state than in $\alpha$.
The energies of these configurations are also i.i.d. Gaussian variables. As a consequence, one can repeat the
previous analysis except that now the number of possible configurations is reduced from $2^N$ to $2^{N(1-c)}$.
Thus the energy distribution reads:
\begin{equation}
{\mathcal N}(E,c)=2^{N(1-c)}\exp(-E^2/(2N))/\sqrt{2\pi N}
\nonumber
\end{equation}
The relationship $-E_{eq}(\beta)/N=\beta$ remains valid but the total entropy is reduced by a factor $-Nc\ln 2$, i.e.
$S(E,c)=S(E,0)-Nc\ln2$, see Fig.2.  At the value $c_K=S(E_{eq},0)/(N\ln2)$ there is an entropy-vanishing transition. 
For $c>c_K$ the system is frozen in the configuration $\alpha$ since all the other ones have an energy
that is higher and determined by $S(E,c)=0$, see Fig. 2. The analysis of this simple model confirms our previous assumptions: 
the configurational entropy per spin is indeed reduced by a factor $-Y c$, here $Y=\beta^2/[2(1-c)]$, and the low-temperature
ideal glass is known: it is the high-temperature configuration $\alpha$.\\
An interesting additional information can be gained making the same analysis on models that 
display also a dynamical glass transition, akin to the Mode-Coupling Theory (MCT) transition. Following \cite{franzmontanari} we expect that since this transition is due to the emergence of metastable states, by freezing particles one blocks unstable relaxation modes
and transforms unstable states in stable ones, thus raising the MCT transition temperature with respect to its $c=0$ value $T_d$.
Within mean field theory this is only possible up to the temperature, $T_o(c)$, where $\Upsilon(c,T)$ vanishes. 
We have analyzed the dynamics and the statics of the spherical p-spin model (details will be presented elsewhere \cite{pspinfreezing}, see also \cite{franzp1}) and found that indeed $T_d$ and $T_K$ are shifted upward by freezing particles. 
The lines $T_{d}(c)$ 
and $T_K(c)$ join at the point at which both transitions disappear, see the phase diagram for the $p=3$ spin model 
in the top panel of Fig.1. This is in agreement with the results found previously by Semerjian and Ricci-Tersenghi in the context of 
combinatorial optimization problems \cite{semerjianricci}.\\
{\bf Critical Properties and Renormalization Group Analysis}.
In order to go beyond mean field theory and obtain the divergence of length and time scales one has to take into account
non-perturbative fluctuations. To this aim we use the approach recently developed in \cite{cbtt} and generalize
it to the case of pinned particles. The analysis is quite cumbersome. In the following we shall only discuss the physical scenario that emerges from our computations; technical details will be presented elsewhere \cite{pspinfreezing}.
Let us first informally review a few important points related to the RFOT theory and the RG analysis of \cite{cbtt}. The order parameter of RFOT is the global overlap $q({\mathcal C}_r,{\mathcal C})$ between a reference equilibrium configuration, ${\mathcal C}_r$, and another equilibrium one, ${\mathcal C}$, that is constrained to be equal to ${\mathcal C}_r$ on the boundary of the sample. The overlap $q$ becomes different from zero\footnote{Depending on the normalization the overlap is zero or assumes a trivial value in the liquid phase.} at the transition because the boundary pinning field forces the configuration ${\mathcal C}$ to freeze in the same ideal glass state to which ${\mathcal C}_r$ belongs. This transition is driven by 
the decrease of the configurational entropy, which plays the role of an ordering field favoring the zero overlap---liquid---phase.
Including, in a renormalization group sense, non-perturbative fluctuations over the length-scale $l$ one finds \cite{cbtt}
that the renormalized liquid is characterized by a configurational entropy $s_c(l)$, proportional to $s_c l^d$, and a term $Y(l)$, akin to a surface tension in the replica field theory formulation, proportional to $Yl^{d-1}$. The latter is related to the surface configurational entropy loss between high and low overlap regions\footnote{This allows us to clarify why we have chosen
similar notations for $\Upsilon$ and $Y$. The reason is that they are related technically, they can be both associated to a surface free-energy cost (but the corresponding free-energy is different) and physically, since the surface free energy mismatch between amorphous states, i.e. $\Upsilon$, can be interpreted as the fundamental origin of the entropy loss due to pinning, i.e. $Y$.}. The values of $s_c$ and $Y$ are obtained from mean field theory. 
By integrating out non perturbative fluctuations until the lengthscale $l$, one finds that a system whose overlap is fixed to one on the boundary of a closed region of size $l\ll l_{PS}$ thermodynamically prefers
to remain in a high overlap state inside this regions, {\it i.e.} amorphous order can be stabilized and the number
of possible amorphous states is $s_c l^d$ below  $l_{PS}$. The contrary happens for 
$l\gg l_{PS}$ since $Y(l)\rightarrow 0$.
Notice that at the transition, when the configurational entropy vanishes,
the point to set length diverges. \\
From the RG point of view, the flow of the coupling constants is the one characteristic of zero-temperature discontinuity fixed points \cite{discfixedpoint}. Indeed,
one can make a useful analogy with the behavior of the ferromagnetic Ising model below $T_c$ in a negative magnetic field, see \cite{krzakala}. The role of the $Y$ and $s_c$ are played respectively by the coupling $J$ and the negative magnetic field $h$: the negative magnetization (low overlap) state is favored by the field but the metastable
one, characterized by positive magnetization (high overlap), can be stabilized on lengthscales smaller than $l_{PS}$ by imposing favorable boundary conditions. Within this analogy the $(T-T_K)\rightarrow 0$ limit corresponds to $h\rightarrow 0$.\\
How does the scenario depicted above changes when one pins particles at random? By doing that, one forces the local overlap to a high value in microscopic regions which are Poisson-distributed inside the sample. Close to all these regions, the system is subjected to an extra field of the order of $Y$, favoring the high-overlap state. This field opposes the global entropic driving force, which instead favors the low overlap state. In consequence, it is easy to grasp
that by increasing the fraction of pinned particles one can induce a transition toward the high overlap state
before the configurational entropy of the unconstrained system vanishes, as found within mean-field theory. The RG analysis becomes more complicated than the 
one for $c=0$ because the local field induced by pinning particles is located in random positions. Because of this quenched
disorder, $Y(l)$ and $s_c(l)$ become random variables and one has to follow the RG flow of their probability law. We have generalized the analysis of \cite{cbtt} to this case and found that the flow is qualitatively identical to the one of the zero-temperature Random Field Ising model, as one would expect on the basis of the analogy discussed above. Indeed, pinning particles translates into fixing a fraction $c$ of randomly located spins in the up position and, hence, leads to an onsite bimodal distribution for the magnetic field, which is negative and equal to $h$ with probability $1-c$ and positive and larger than $2dJ$ with probability $c$ (a field larger than $2dJ$ fixes the spin in the up position on a hyper-cubic lattice in $d$ dimensions).\\
Let us first describe and discuss our results for temperatures slightly above $T_K$. In this case, by increasing $c$ toward its critical value we find results very similar to the ones obtained approaching $T_K$ at $c=0$: the average value of the surface tension
on the lengthscale $l$ increases as $\overline Y l^{d-1}$, whereas the average value of the configurational entropy
increases as $k(c_K(T)-c)l^d\propto \overline s_c(c,T)l^d$. The corresponding variances become much smaller than the averages, so that the random variables $Y(l)$ and $s_c(l)$ almost do not fluctuate around their means. $\overline Y$ and $k$ depend on $T$ and $c$ and are of the order of one, whereas $\overline s_c(c,T)$ is the average configurational entropy density for a given value of $c$ and $T$; it goes to zero linearly at $c_K(T)$. The length for which 
$Y(l)$ becomes of the order of $s_c(l)$ is the point-to-set length, $l_{PS}(c,T)$, for the pinned system 
(which is much larger than the one for the unconstrained liquid). For $l\gg l_{PS}$ one finds that $Y(l)\rightarrow 0$ and hence one cannot stabilize amorphous states.
The scaling of $l_{PS}$ with $(c_K(T)-c)$ is the usual one:
\[
l_{PS}\propto \frac{\overline Y}{\overline s_c(c,T)}\propto \frac{\overline Y}{(c_K(T)-c)}\, .
\]
As already discussed in \cite{cbtt}, the physical picture resulting from the RG analysis is that the renormalized system on the point-to-set lengthscale is like a liquid at its ``onset temperature''. The difference with a normal liquid are the typical value of the free energy contributions, which respectively are of the order of  $\overline Y l_{PS}^{d-1}$ and $\overline s_c l_{PS}^d$, both very large compared to the temperature. This suggests that the typical relaxation time follows an Arrhenius law and, therefore, a generalized Adam-Gibbs relation \cite{AG}:
\begin{equation}\label{arrhenius2}
\log \tau \propto \frac{ \overline Y l_{PS}^{d-1}}{T} \propto \frac{ \overline Y^d }{T\overline s_c^{d-1}}\propto
\frac{1}{(c_k-c)^{d-1}} \ .
\nonumber
\end{equation}
Important differences with the $c=0$, $T=T_K$ case emerge when varying $c$ at higher temperatures above $T_K$. Physically,
the main consequence of fixing a higher reference temperature is to increase the value of the configurational entropy 
of the unconstrained liquid. Hence,
the fraction of pinned particles needed to counterbalance this effect and to induce the transition increases, see Fig.1.
By doing so the variance of the random field, proportional to $c$ (for small $c$),  increases too.
We find that above a certain temperature, $T_h$, disorder fluctuations induced by pinned particles are so strong enough to make the glass transition disappear, in agreement with what was found for the RFIM when increasing the disorder \cite{middleton,aharony}.  We studied the RG flow approaching $T_h$. We found that the glass transition line ends in a second order transition point and that the long-distance physics is dominated by the RFIM zero-temperature critical fixed point. This could be expected 
\cite{franzp2} on the basis of the results \cite{franzparisiriccirizzo}. 
We found that the means and the variances of $Y(l)$ and $s_c(l)$ scale as $l^{\theta}$, with $\theta=1.491(4)$ at $T=T_h$ and $c=c_K(T_h)$ (to be compared to the one obtained for the RFIM by simulations $\theta=1.49(3)$\cite{middleton}). When $T$ is close, but smaller than $T_h$, the critical properties of the glass transition become quite complicated since the RG flow is first attracted by the RFIM zero temperature fixed point and
approaches the standard first-order discontinuity fixed point for very large lengthscales only. Calling $\nu$ the correlation length exponent for the RFIM (and using the standard notation for the other exponents \cite{middleton,aharony}), we find that only for $l_{PS}\gg 1/(T_h-T)^\nu$ the usual scaling laws discussed above are obtained. Instead, for  $l_{PS}\ll 1/(T_h-T)^\nu$, {\it i.e.} for $(T_h-T)^{\nu (d+\overline \eta -2\eta)/2}\ll c_k-c\ll O(1) $, one finds:
\[
l_{PS}\propto \frac{1}{(c_k-c)^{2/(d+\overline \eta -2\eta)}} \quad , \quad\log \tau \propto  \frac{1}{(c_K(T)-c)^{2\theta/(d+\overline \eta -2\eta)}}
\]
Because in d=3 the exponent $\theta$ is very close to $1.5$ and $\overline \eta \simeq 2\eta$, the relaxation time is expected to grow in this regime almost according to the so called Vogel-Fulcher-Tamman law but with respect to the variable $c_K(T)-c$ instead of $T-T_K$, {\it i.e} $\log \tau \propto (c_K(T)-c)^{-1}$.\\
A final important outcome of the RG analysis concerns the $T-c$ phase diagram. 
In particular we find that the phenomenological law $c_K(T)\propto s_c(T)/Y(T)$ is quite well obeyed for $T<T_h$
and that the temperature $T_h$ corresponding to the endpoint of the critical line is depressed with respect to its mean field value $T_o$. The phase diagram in the lower panel\footnote{We cannot draw the equivalent of the MCT line because this is defined within mean field theory only. The value of $T_d$ (at $c=0$) that we use and draw in the lower panel of Fig. 1 is obtained from mean-field theory.} of Fig.1 corresponds to $T_d/T_K\simeq 1.4$, a value typical of fragile liquids like OTP \cite{dzero} ($T_d^{OTP}\simeq285K$, $T_K^{OTP}=203K$). In this case $T_h$ is very close to $T_d$: $T_h\simeq 280K$. Within our approximate treatment  the decrease of $T_h$ with respect to $T_o$ depends on the ratio of the mode coupling temperature to the Kauzmann temperature. Smaller values of $T_d/T_K$ correspond to more fragile supercooled liquids and lead to an increase of $T_h/T_d$, that can reach values larger than one (and viceversa for more strong liquids). \\
{\bf Other Ways of Freezing Particles and the Role of Disorder}. 
Pinning particles at random in space introduces a lot of quenched disorder and, hence, leads to a substantial decrease
of the endpoint of the $c_K(T)$ line. In order to reduce this effect,  
one can pin the particles closest to an ordered template, {\it e.g.} a periodic one \cite{bktobe}. 
This leaves as source of quenched disorder, the fluctuations of the reference equilibrium configuration only \cite{franzparisiriccirizzo,wolynesrfim}. In this case we expect similar results, the only change
being the precise form of $c_K(T)$ and the value of $T_h$, which is expected to be higher. Actually, 
the relationship with the RFIM 
Another case considered in the literature \cite{bk} corresponds to freezing
all particles outside two walls at distance $\ell$. Following the Kac analysis of Franz \cite{franzoupchapter},
one finds qualitatively the mean field phase diagram of the top panel of Fig. 1, where now $\ell$ is the control parameter playing the role of $c^{-1}$. The transition takes place at a value of $\ell$ proportional to the point to set length of the 
unconstrained system\footnote{Instead, for randomly pinned particles, the distance between frozen particles
at the transition scales as the point to set length at a power less than one, see the phenomenological arguments introduced before and \cite{lengths}.}. 
The effect of disorder is not taken into account in the Kac analysis. Since 
the positions of the particles pinned beyond the walls are random, the constraint on the coarse-grained overlap is not
homogeneous close to the walls and lead to an effective quenched disorder. 
Applying our RG treatment to the two wall case and including the effect of disorder, 
one would find results similar to the previous ones but with $d\rightarrow d-1$. This has two important consequences.  
First, as for randomly frozen particles, the mean field phase diagram is modified: the endpoint of the critical 
glass transition line takes place at a temperature less than the onset temperature computed by mean field
or Kac-like techniques. Second, because the system inside the two walls is $d-1$ dimensional, the lower critical dimension
is higher than the one expected in the case of randomly frozen particles. No first order transition takes place in two dimensions for the Ising model in presence of quenched randomness \cite{aizenman}. Since within our framework glass transitions are related to the first order phase transitions of the RFIM, we conclude that glass transitions can be only induced in systems with dimensions equal to four or
higher in the case of two walls. Instead, in the case of randomly pinned particles, also three dimensional systems display glass transitions when $c$ is increased. \\
The final case we shall discuss corresponds to particles that are frozen not from an equilibrium configuration but instead completely at random, {\it i.e.} from an infinite-temperature configuration. This is akin to studying liquids in 
presence of external quenched randomness \cite{dasgupta}, a case relevant for porous media \cite{krakoviack}. 
Because the quenched disorder is not tailored to an equilibrium configuration at temperature T, we do expect qualitative changes. In fact the phase diagram should be qualitatively similar to those obtained in \cite{dasgupta,krakoviack}, which are quite different from ours and resemble that of the p-spin spherical model
in an external magnetic field: above a certain fraction $c$ the glass transition becomes continuous (the Edwards-Anderson parameter has no discontinuous jump) but does not disappear; moreover, the glass phase for $c>c_K$ cannot be reached from high temperature without crossing a transition line, contrary to what is shown in Fig. 1.
We shall discuss our phase diagram in more detail in the following. \\
{\bf Glass Transitions Induced by Random Freezing: Novelties, Consequences and Numerical Verifications}. 
The most striking feature of the phase diagram shown in Fig. 1 is that it is possible to follow a path that connects the liquid and the glass phase without crossing any phase transition, as for 
the liquid-gas phase transition\footnote{At variance with the liquid-gas case, changing temperature in the phase diagrams of Fig. 1 does not simply mean changing a control parameter. Different temperatures lead to different configurations of frozen particles, hence changing temperature means studying different systems. 
} (see \cite{pspinfreezing} for more details). Afterall,
this is not so surprising since the ideal glass phase coincides up to small fluctuations with 
the initial configuration from which one freezes particles. Thus, one does not have to
equilibrate the system on impossibly large time-scales  or to solve a daunting optimization problem in order to reach
the ideal glass. This should allow one to perform much more detailed investigations of the glass transition. In particular, the belief that it is impossible to prove (or disprove) the existence of the ideal glass transition because the time-scale for equilibration diverges so rapidly maybe ill-founded. For instance, the RFIM
also has a transition with activated dynamic scaling and despite numerical difficulties several properties of the transition have been obtained by numerical simulations. 
The main difference between the transition of unconstrained supercooled liquids and the RFIM's one
is that for the former the low temperature phase---the ideal glass---is not known.
 This prevents one from approaching the glass transition from both ends, from devising clever algorithms to equilibrate the systems, etc. In the case of frozen particles, this is no more the case. 
Thus, one can endeavor to demonstrate the existence 
of the ideal glass transition by applying standard statistical physics techniques.
Ideally, one could first study the transition at a finite temperature well above $T_K$, at a finite $c_K(T)$. 
This would be already a major result in itself. 
Afterwards, by decreasing the temperature, one could measure the $c_K(T)$ line and show, at least by extrapolations, 
the existence of the ideal glass transition for usual, unconstrained, supercooled liquids.  
Even though not impossible, numerical verifications will be very challenging (work is underway \cite{bktobe}). We think that one should start first
by studying finite-size scaling.
 If there is indeed a transition, the averaged probability distribution function of the overlap with the reference configuration should behave as in first order phase transitions: a single peak at small (large) overlap below (above) $c_K$ in the thermodynamic limit. Only at $c_K$, a two peak structure should be present. The corresponding finite size scaling is expected to be similar to the 
one of first order phase transitions in presence of quenched randomness, see e.g. \cite{fisher}.
The effect of finite size on the relaxation time $\tau$ is instead different. Actually, already the thermodynamic 
limiting behavior is different: $\tau$, measured for instance by the intermediate scattering function, diverges approaching the critical line from the liquid side but not from the glass side\footnote{This notwithstanding, the self and the collective relaxation times can be both very large, although finite on the glass side.}. Finite sizes make $\tau$
finite at the transition. Coming from the glass side $\tau$ should increase very fast on a window that shrinks to zero in the
thermodynamic limit. \\
Actually, an instructive case to grasp 
the main ideas and difficulties is provided by the three dimensional Ising model below $T_c$. Take an equilibrium configuration in presence of
a positive magnetic field $h$, freeze a fraction $c$ of spins at random and revert the field to $-h$. Now the original 
configuration is metastable for $c$ smaller than a critical value $c^*$, and stable for larger ones. Despite some important differences, the situation 
is very reminiscent of the one we discussed for supercooled liquids. Indeed, our RG analysis is precisely related to this problem at zero temperature
with $h$ playing the role of the configurational entropy, the positive and negative magnetization playing the role of 
high and low overlap. The life time of the metastable state, which diverges for $c^*-c \rightarrow 0$,  is analogous, {\it mutatis mutandis}, to the relaxation time that takes to the liquid to decorrelate from the initial configuration and 
that diverges for
$c_K-c \rightarrow 0$. 
We have studied the behavior of the overlap with the reference
configuration as a function of time, see Fig. 3. 
We also explicitely verified that the overlap distribution  
behaves in the way described above, see the inset of Fig. 3. 
We found that by increasing the value of $c$ the system spends more and 
more time in the metastable high-overlap state before equilibrating at low overlap. The timescale for relaxation at low overlap diverges
at a critical value $c^*(T,h)$ with the same law discussed before in the glass transition case
\footnote{The major important difference between this simple Ising model and real liquids is that the original reference configuration is not an equilibrium one. The main consequence is that the system is not automatically at equilibrium after having frozen out particles; thus, relaxation and equilibration time are not the same, contrary to the case of super-cooled liquids.}:  
$\log \tau \propto 1/(c-c^*)^{d-1} $. 
The curves in Fig.3 reproduce qualitatively well the ones obtained in supercooled
liquids by Berthier and Kob \cite{bk} (besides the trivial difference that asymptotic low overlap value is negative).
This is an indication that the RFOT theory, and the related physical mechanism we have unveiled in this work,  
provide a viable explanation for the mysterious and generic dramatic 
slowing down of the dynamics, which has been observed in simulation when pinning particles of super-cooled liquids \cite{bk,Kim1,Kim2,Procaccia}. In order to show that this slowing down is indeed related to a phase transition, one needs
to study finite size scaling as discussed above.\\
\vspace{-0.8cm}
\begin{figure}[h]
\hspace{-0.35cm}
\includegraphics[width=0.45\textwidth]{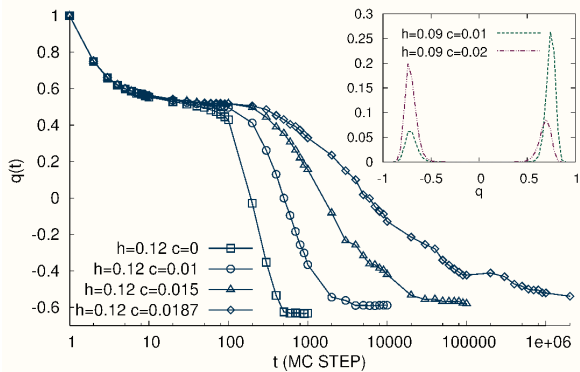}\\
\vspace{-0.2cm}
\caption{Overlap with the reference configuration as a function of time for the 3D Ising model at $T=4$, $h=0.12$ on a cubic lattice with $N=8*10^3$ spins ($J$ is set equal to one). Increasing the fraction of frozen spins it takes a much larger time to escape from
the high overlap phase. Inset: overlap distribution slightly above and below $c^*$  ($T=4$, $h=0.09$, $N=10^3$).\vspace{-1.3cm}\\}
\label{fig3}
\end{figure} 
$\phantom{a}$\\
$\phantom{a}$\\
{\bf Conclusions}.
In this work, motivated by the numerical results \cite{bk,Kim1,Kim2,Procaccia} we have studied the effect of pinning particles of an equilibrated supercooled liquid configuration. We have focused on geometries such that the number of unfrozen particles is infinite. In these settings, pinning can lead to a true phase transition and not just a cross-over, like
in the cavity case studied previously \cite{bb,bbcgv}. Indeed, we have shown that pinning leads to a decrease of the configurational entropy and, within the RFOT theory, to an ideal glass transition when the confinement parameter $c$ is increased, even for relatively high temperatures, e.g. close to the Mode-Coupling transition temperature. The relaxation timescale\footnote{Only the relaxation time of collective (as opposed to self) correlation functions diverge.} is infinite along the $c_K(T)$ line
and finite everywhere else, in particular for $c>c_K(T)$. Although in the latter regime motion of single particles
can be quite slow, as it is in a crystal, we predict that if properly equilibrated the system should show a nonmonotonous
behavior of the relaxation time as a function of $c$ for $T<T_h$.  \\
Pinning particles may be expected to generically slow down the dynamics whatever is the correct underlying theory
(RFOT theory, facilitation theories \cite{gcrev}, frustration limited domains \cite{gilles}, etc.).
However, the nonmonotonous behavior of the relaxation time and the concomitant thermodynamic phase transition are expected within RFOT only\footnote{And also within phenomenological approaches related to it like the Adam-Gibbs one. From this point of view our results rationalize the ones of \cite{Procaccia} based on the Adam-Gibbs law.}, thus opening the way to new and crucial tests of glass transition theories. For example, in the case
of kinetically constrained models (KCM), we expect that one can induce dynamical glass transitions by increasing $c$ above a critical value $c_d$.
However, since the kinetic constraint can be violated in reality, the relaxation of super-cooled liquids would proceed via alternative and very slow relaxation mechanisms for $c>c_d$ and would lead to a behavior very different from the one expected within RFOT.  A detailed study of the effect of pinning particles in  Kinetically Constrained Models would be worthwhile.\\
We notice that our approach shares important similarities with two other approaches: the one of Franz and Parisi \cite{fpepsilon}, consisting in biasing the 
thermodynamics by favoring configurations having a high overlap with an equilibrium one; and the one
of Garrahan and Chandler \cite{GCspacetime}, consisting in biasing the dynamics by favoring dynamical trajectories with low dynamical activity. The mean-field result analysis of the former leads to a phase diagram that is indeed analogous to ours (the bias strength plays the role of $c$). We expect this relationship to persist even beyond 
mean field. Some numerical simulations of three dimensional liquids provide promising results in favor of this approach
 \cite{fpepsilon,chiaranuc}. Theoretical analysis of KCMs and numerical simulation of glass-forming liquids \cite{GCspacetime,vanwijland} show that also in Garrahan and Chandler approach one finds a phase diagram similar to the one in Fig.1.    
There are however important fundamental differences in this case. 
The main one is that the phase transition taking place 
at $c_K(T)$ and shown in Fig.1 exists within the RFOT theory only, contrary to the dynamical transition studied in \cite{GCspacetime,vanwijland} that takes place in facilitation-based models but also in
those described by the RFOT theory (see \cite{jackgarrahan}).\\
We end this paper by stressing again that one of the main
advantages of the glass transition induced by random freezing is that the ideal glass phase can be obtained easily. 
Hence, if a transition indeed takes place at $c_K$, it can be very likely  studied thoroughly, in particular by finite-size 
scaling. Since we find that the critical properties for $c_K-c\rightarrow 0^+$ and $T>T_K$ are the same ones as those of the ideal 
glass transition when $T\rightarrow T_K$ for $c=0$, studying the effect of freezing particles appears to be a new and promising way to understand the glass transition of supercooled liquids.  These studies can be performed in supercooled liquids by 
numerical simulations and also on colloids by experiments using optical trapping.   
\vspace{-0.4cm}
\begin{acknowledgments}
We thank L. Berthier and W. Kob for letting us know their results in advance and for discussions.  
We thank J.-P. Bouchaud and G. Tarjus for suggestions and for very helpful comments on 
the first version of this manuscript and A. Cavagna, S. Franz, T. Grigera, V. Krakoviack, G. Parisi, D.R. Reichman, G. Semerjian, G. Tarjus, M. Tarzia, F. Zamponi  for comments and discussions.
\end{acknowledgments}

\end{article}

\end{document}